# The Fundamentals of Generalized Electrodynamics

Tomilin A.K.[*]

*D. Serikbaev East-Kazakhstan State Technical University,
Ust-Kamenogorsk, Kazakhstan*

**Abstract**

The article presents an alternative approach to the definition of vector electrodynamic potential and its properties. It is shown that generally it has vortical and potential components. The system of differential equations of generalized electrodynamics (the macroscopical theory) is written down; and on the basis of these equations the mechanism of occurrence of the electromagnetic waves spreading in the direction of vector $\vec{E}$ (longitudinal *E*-waves) is explained. It is demonstrated that the new macro-scopical theory is in accordance with quantum electrodynamics. The author investigates the issue of physical pithiness of longitudinal electromagnetic *E*-waves and specifies the references to the experimental researches confirming the new theory.

---

[*] Electronic address: tomilin@ukg.kz

## 1. Introduction

As it is known, modern magnetostatics is based on the following differential equations:

$$\vec{H} = \frac{1}{\mu_0} rot\vec{A}, \qquad (1)$$

$$div\vec{A} = 0, \qquad (2)$$

where $\vec{A}$ - vector potential, $\vec{H}$ - tension of the magnetic field, $\mu_0$ - magnetic constant.

In textbooks on electrodynamics it is usually said that vector potential has no physical meaning and is used as auxiliary function but condition (2) is entered to disambigue this function. According to condition (2), the lines of the vector must be locked-in, and it means that the field of this vector is rotational.

As it is known, the vector potential satisfies the equation of Poisson. We shall calculate the vector potential of the field of infinitely long rectilinear current $J$, directed along axis z:

$$\vec{A}(x',y',z') = \frac{\mu_0 J}{2\pi} \int_0^\infty \frac{dz}{\sqrt{x'^2 + y'^2 + (z-z')^2}} \vec{z}^{\,0} =$$

$$= \frac{\mu_0 J}{2\pi} ln \left| (z-z') + \sqrt{x'^2 + y'^2 + (z-z')^2} \right|_0^\infty \vec{z}^{\,0}. \qquad (3)$$

When computing this formula it is necessary to define the value of $\vec{A}$ when $(z-z') \to \infty$, that is to say the condition of vector potential is required. As such condition the following is taken:

$$\vec{A}_\infty = 0. \qquad (4)$$

Then

$$\vec{A}(x',y') = -\frac{\mu_0 J}{2\pi} ln \left| \sqrt{x'^2 + y'^2} \right| \cdot \vec{z}^{\,0}. \qquad (5)$$

It is not difficult to show that divergency of the vector expressed by formula (5) equals zero, so condition (2) is executed in this instance. Condition (4) has provided locking-in of vector $\vec{A}$ lines, but execution of the condition (2) is due to the use of condition (4). After calculating $rot\vec{A}$, we shall get the known formula for induction of the magnetic field, created by rectilinear endless current.

The argument of the logarithm in formula (5) changes from zero ad infinitum, at the same time the sign of the function also changes: if the value of the argument is less than one, function (5) is positive, but if value of the argument is greater than one, it is negative. Consequently, vector $\vec{A}$ is directed on current if it is near a conductor, and it is directed against the current if it is far from a conductor. Locking-in of vector $\vec{A}$ lines occurs at infinity, that confirms fairness of the condition (4). According to (1) the rotational field of vector $\vec{A}$ generates vector magnetic field $\vec{B}$.

At determination of the magnetic field of separate locked-in current sidebar, the characteristics of vector potential turn out to be the same, i. e. corresponding to equations (1) and (2). Usually the study of vector potential characteristics is limited to the above.

However the theory built on such basis, does not explain many electrodynamic phenomena and results of experiments that are told about in publications [3-9].

## 2. Starting positions and consequences

We shall calculate the vector potential of the magnetic field created in free point $M(x',y',z')$ by rectilinear current $J$, flowing along the conductor of defined length. Certainly, we get principle question about justification of such statement of the problem. Usually it is considered that constant currents are locked-in, and on this base the currents of defined length, that are not locked-in, are not considered. However, if we present the locked-in current sidebar as electromechanic system, there arises the question about the interaction between its parts at the account of internal powers. Such approach requires consideration of separate current lengths, and interactions between them by means of the fields they create.

If we link beginning of coordinate system with one of the ends of the current segment, axis $z$ we shall direct along the current (fig. 1), so we shall get the following:

$$\vec{A}(x',y',z') = \frac{\mu_0 J}{4\pi} \ln\left|\frac{L - z' + \sqrt{x'^2 + y'^2 + (L-z')^2}}{\sqrt{x'^2 + y'^2 + z'^2} - z'}\right| \cdot \vec{z}^0. \tag{6}$$

We notice that it was not necessary to enter any conditions. Mark the positive values:

$$r_1 = \sqrt{x'^2 + y'^2 + z'^2}; \quad r_2 = \sqrt{x'^2 + y'^2 + (L-z')^2}. \tag{7}$$

They are the modules of radius-vectors, called on to the point $M(x',y',z')$ from beginning and end of the current segment accordingly. Comparing the numerator and denominator of the expression under the logarithm sign in (6), it is not difficult to see that

$$\frac{L - z' + \sqrt{x'^2 + y'^2 + (L-z')^2}}{\sqrt{x'^2 + y'^2 + z'^2} - z'} > 1,$$

because the sum of lengths of the two sides of the triangle showed in fig. 1 is always more than its third side:

$$L + \sqrt{x'^2 + y'^2 + (L-z')^2} > \sqrt{x'^2 + y'^2 + z'^2},$$

or

$$L + r_2 > r_1.$$

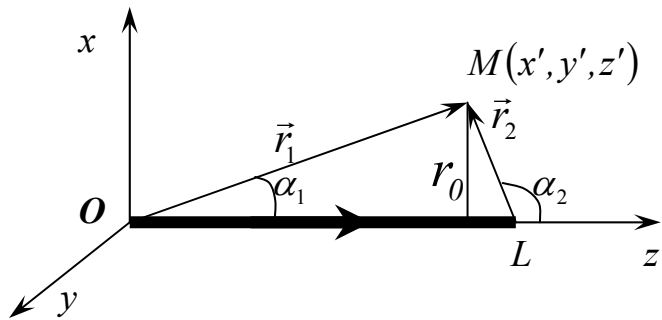

Fig. 1

So, according to (6), the lines of vector potential have one direction in this case. Therefore there should exist the sources and drains of the field of vector $\vec{A}$.

In publications of Nikolaev G.V. [5-6] the hypothesis of magnetic field component existence, which is not taken into account in Maxwell's electrodynamics, is proved. Really, the identification of a magnetic field with the help of iron sawdust, arisen in the earliest stage of studying magnetism, is not reasoned. Is it possible to describe electromagnetic interaction in all cases using only knowledge about magnetic field lines? Such question was not put in due time. It also has resulted in limited nature of modern electrodynamics.

Austrian professor S. Marinov from reasons of the basis of field theory has offered [7-8] to determine $H^*$ function through vector potential as follows:

$$H^* = -\frac{1}{\mu_0} div\vec{A}, \qquad (8)$$

where $H^*(x', y', z')$ - scalar function. At the suggestion of Nikolaev G.V., the concept *of scalar magnetic field (SMF)* which is described by this function is introduced. Thus, it is suggested to use two functions for the description of full magnetostatics field: vector $\vec{H}(x', y', z')$ and scalar $H^*(x', y', z')$.

Indeed according to *the basic theorem of the field theory (Stox-Helmholz)*: if divergency and rotor of a field (in this case $\vec{A}$), turning to zero at infinity, are determined in each point $\vec{r}$ of some area, the field of vector $\vec{A}(\vec{r})$ may be presented as a sum of potential and solenoidal fields everywhere in this area. Thus, the correlation (8) cancels artificial calibration (2), and allows to work out generalized magnetostatics.

It is clear that the vector potential $\vec{A}$ introduced this way has other properties, than in usual electrodynamics. First of all, from (8) it follows that the field of vector $\vec{A}$ has sources and drains which are in SMF. Sources of the field of vector $\vec{A}$ are in negative SMF, and drains are in positive SMF.

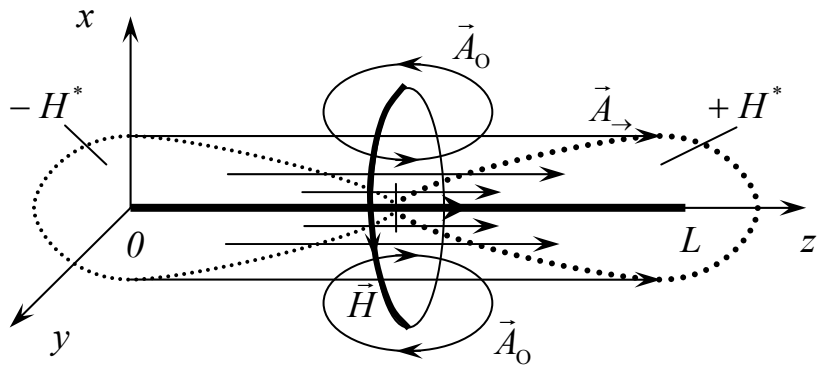

Fig. 2

Fig. 2 conventionally represents the field of vector potential $\vec{A}$, Including rotational $\vec{A}_\circlearrowleft$ and potential $\vec{A}_\rightarrow$ components, and also vector magnetic field $\vec{H}$ and scalar magnetic field $H^*$, created by rectilinear current segment of certain length.

As a result of calculation of the function divergency (6), we have:

$$div\vec{A} = \frac{\partial A_z}{\partial z'} = \frac{\mu_0 J}{4\pi}\frac{(r_2 - r_1)}{r_1 r_2}. \qquad (9)$$

As at any values of $r_1$ and $r_2$, size of $div\vec{A} \neq 0$, that SMF in this case is created:

$$H^*(x',y',z') = \frac{J}{4\pi} \frac{(r_1 - r_2)}{r_1 r_2} = \frac{J}{4\pi r_0}(\sin\alpha_1 - \sin\alpha_2), \tag{10}$$

We research function (10). Fig. 3 shows the diagram of dependence $H^*(0,0,z')$, that is this function is determined in the points laying on axis z, thus $r_1 = |z'|$, $r_2 = |L - z'|$. Apparently from the diagram, function $H^*(0,0,z')$ is sign-variable and on the ends of current segment *AB* has disruptions. Distribution of function $H^*(0,0,z')$ corresponds to SMF presented in fig. 2. It is noticeable that SMF gradient $gradH^*(0,0,z')$ appears along a conductor, it is directed along the current flowing in it. It is possible to formulate the general rule: *if to look from the midpoint of the segment in the direction of the current flowing in it, positive SMF is created ahead, and negative SMF is created behind.*

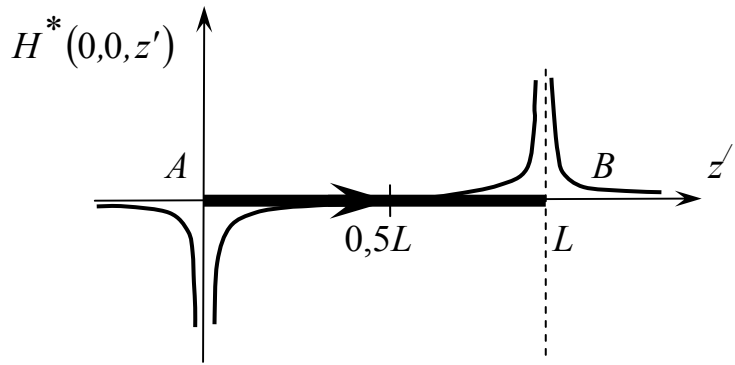

Fig.3

The above said leads to an important conclusion: *the vector potential generally has vortical and potential components:*

$$\vec{A} = \vec{A}_\circlearrowleft + \vec{A}_\rightarrow. \tag{11}$$

Thus formulas (1) and (8) can be put accordingly as follows:

$$\vec{B} = rot\vec{A} = rot\vec{A}_\circlearrowleft, \tag{12}$$

$$B^* = -div\vec{A} = -div\vec{A}_\rightarrow. \tag{13}$$

Taking into account substance properties, the connection between intensity $H^*$ and induction $B^*$ of SMF is represented by the following correlation:

$$B^* = \mu'\mu_0 H^*. \tag{14}$$

Concerning dimensionality of SMF characteristics the full analogy to the corresponding characteristics of a vector magnetic field is observed: $H^*$ is measured in *A/m*, $B^*$ is measured in *Tl*.

In his monographies Nikolaev G.V. [5-6] gives the equations which are suggested as fundamental to generalized magnetostatics:

$$div\vec{H} = 0, \tag{15}$$

$$rot\vec{H} + gradH^* = \vec{j}. \tag{16}$$

It is follows from the equation (16) that the current of conductivity derivates also a scalar (potential) magnetic field, except for a usual vector (solenoidal) magnetic field. Let's pay attention to the fact that equation (16) also corresponds to Stox-Helmholz theorem with reference to a field of currents $\vec{j}(\vec{r})$.

Having substituted in (16) equations (1) and (8), we shall receive:

$$rotrot\vec{A} - graddiv\vec{A} = \mu'\mu_0 \vec{j}.$$

In the result we have equation of Poisson:

$$\Delta \vec{A} = -\mu'\mu_0 \vec{j}. \tag{17}$$

Thus, the vector potential at such approach, as well as in traditional magnetostatics, satisfies the equation of Poisson, however, at its computing the condition (2) was not required.

As one of the most important conclusions at this investigation phase we shall note, that both components of a magnetic field (vector and scalar) are determined by means of vector electrodynamic potential. Consequently, ***vector electrodynamic potential is necessary to recognize as the basic characteristic of full magnetostatic field.***

The question about conditions of creation SMF by complete current system seems very important. At its solution it is necessary to rely on the properties of vector potential [3-4]. Let's consider rectangular current contour. In any point of space there are imposed magnetic fields from four current segments forming it. In any point of space, except for the point of diagonal crossing of rectangular $O$, vector $\vec{A}$ is other than zero. The configuration of the vector magnetic field created by the rectangular contour is known.

Let's choose any point of space and make in it radius-vectors from all four corners, having designated them accordingly $r_1, r_2, r_3, r_4$ (fig. 4).

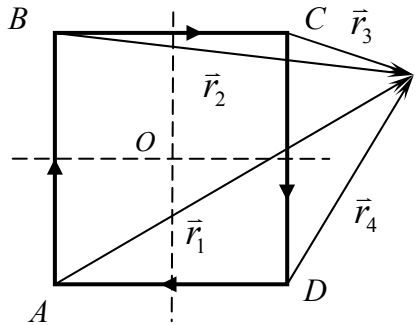

Fig.4

It is not difficult to see, that

$$H^*(x',y',z') = \frac{J}{4\pi}\left(\frac{r_1 - r_2}{r_1 r_2} + \frac{r_2 - r_3}{r_2 r_3} + \frac{r_3 - r_4}{r_3 r_4} + \frac{r_4 - r_1}{r_4 r_1}\right) = 0.$$

Thus, SMF is not created by rectangular locked-in current contour in any points of space. Really, for the locked-in contour $div\vec{A} = 0$, therefore $H^* = 0$. Obviously, it is possible to make a more general conclusion: ***locked-in current does not create SMF at any form of a contour in which it flows.*** This conclusion agrees to the traditional theory.

Now we shall consider the system of two identical rectangular current contours located in one plane (fig. 5). It is clear that in any point of space, chosen accidentally, the components from all eight current segments are imposed. It is not difficult to show, that in the point $M$, located on the axis of symmetry of electric system $Ox$, the vector potentials of current segments $AD$ and $D_1A_1$ and also $CB$ and $C_1B_1$ are in pairs compensated. The vector potentials of currents $AB$ and $A_1B_1$ are directed to the right side, and currents of $CD$ and $C_1D_1$ – to the left. Let's make it a point that the sum of the two first potentials, undoubtedly, is more than the sum of two second ones. It is obvious, that the resulting vector potential in point $M$ is other than zero, and it is directed to the right. It is possible to show that on axis $Ox$ $rot\vec{A}=0$, consequently, the vector magnetic field on the axis of symmetry is missing. It is easy to verify that after considering the picture of the magnetic field lines modelling a usual magnetic field.

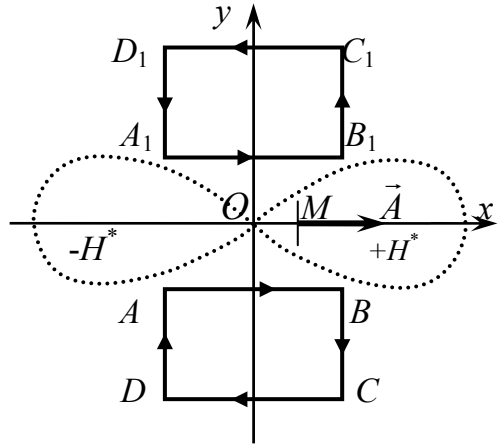

Fig. 5

Having calculated the divergence of vector potential, we shall receive expression for intensity of SMF in any point on axis $Ox$:

$$H^*(x',0)=\frac{J}{2\pi}\left[\frac{(r_1-r_2)}{r_1 r_2}+\frac{(r_3-r_4)}{r_3 r_4}\right], \qquad (18)$$

where $r_1, r_2$ - modules of radius-vectors which have been carried out accordingly from points $A$ (or $A_1$) and $B$ (or $B_1$) to point $M$. $r_3, r_4$ - modules of radius-vectors drawn accordingly from points $C$ (or $C_1$) and $D$ (or $D_1$) to point $M$. We investigate function (18). In the centre of the current system symmetry in point $O$ there is no SMF:

$$H^*(0,0)=0.$$

It is very important in examining any current system to determine a special point in which SMF is equal to zero. Knowing the direction of vector potential $\vec{A}$ in relation to this point, it is possible to determine the sign of SMF, according to the rule: ***SMF has a positive sign where vector potential $\vec{A}$ is directed away from the special point, and a negative sign where vector $\vec{A}$ is directed towards the special point.*** Having applied this rule it is easy to see that in the right part of the gap between the electric contours SMF has a positive sign, while in left – negative. These areas are represented in fig. 5.

Once again we shall emphasize, that the initial characteristic of the full magnetic field created by a current system, is the vector potential. Such approach allows to avoid the erroneous decision: if the scalar magnetic field is not created by a separate contour so the resulting field from two contours is also equal to zero.

The contours represented in fig. 5 are formed in diametrical section of a toroid. The toroid represents the ideal electrostatic system creating a magnetic field in which a vector and scalar components are positionally divided: the vector field is completely inside the toroid, and scalar - outside (fig. 6). The intensity of SMF on an axis of a toroid is determined by the following formula:

$$H^*(x',0) = n\frac{J}{2\pi}\left[\frac{(r_1 - r_2)}{r_1 r_2} + \frac{(r_3 - r_4)}{r_3 r_4}\right], \tag{19}$$

where $n$- quantity of pair coils of a toroid.

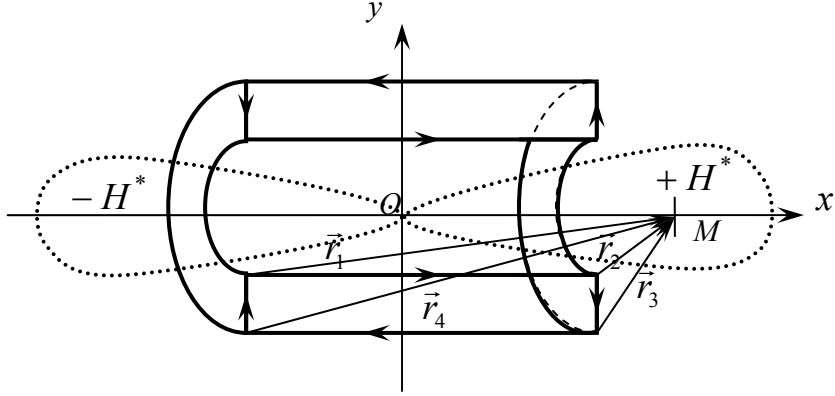

Fig. 6

If to accept the height of a toroid as equal to $2h$, and to designate its internal and external radiuses accordingly $r_T, R_T$, it is convenient to present the segments which are included in the formula (19) as:

$$r_1 = \sqrt{r_T^2 + (x+h)^2}, \quad r_2 = \sqrt{r_T^2 + (x-h)^2}, \quad r_3 = \sqrt{R_T^2 + (x-h)^2}, \quad r_4 = \sqrt{R_T^2 + (x+h)^2}.$$

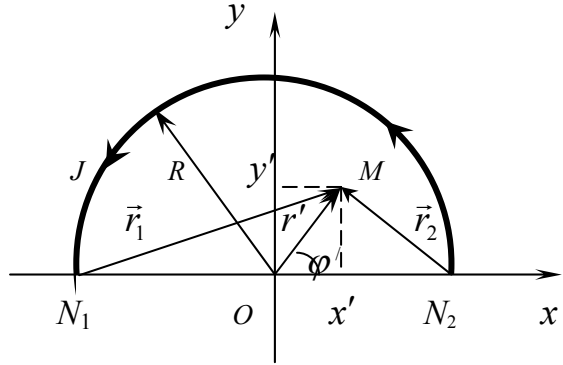

Fig. 7

It is not difficult to show, that the locked-in ring current does not create SMF. Vector potential of semiannular current with radius $R$ (fig. 7) is calculated by the formula (10) as it is possible to lock-in a contour by rectilinear segment of $2R$ length. Thus it is convenient to use polar coordinates, then in (10):

$$r_1 = \sqrt{R^2 + r'^2 + 2Rr'\cos\varphi'}, \quad r_2 = \sqrt{R^2 + r'^2 - 2Rr'\cos\varphi'}.$$

SMF may be created by constant magnets. For the first time it was noticed *by Nikolaev G.V.* who has created a special magnet. Magnet of Nikolaev (MN) represents a cylindrical magnet sawn on diameter in two parts which are turned against each other half-way (180 degrees) (fig. 8). Magnetic field created by such magnet is modelled by semiannular and radial currents. The theory developed above allows to represent SMF in this figure taking into consideration function signs. From considerations of symmetry it is not difficult to guess that there is a special point $O$ in the centre of the current system. In the left part the vector potential is directed to point $O$, consequently, SMF has the negative sign here, and in the right part vector $\vec{A}$ is directed from point *O*, consequently, positive SMF is created here.

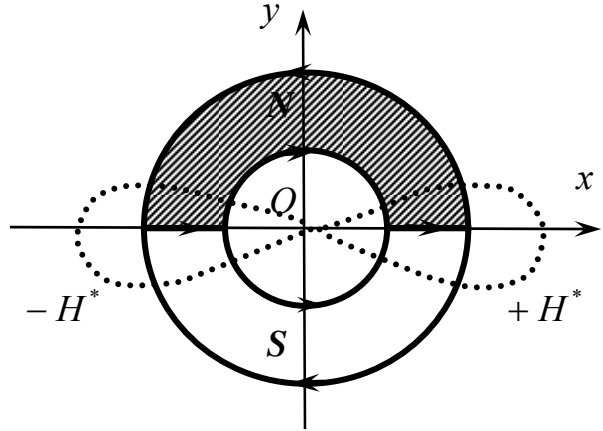

Fig. 8

For creation of SMF it is possible to use a pair of flat magnets connected with their lateral sides. Thus magnetostatic system is formed, which is modelled by the current contours shown in fig. 5.

Existence of SMF is proved by numerous experiments, in particular about using a magnet of Nikolaev [5-6, 9]. It is shown, that the force directed along a current or against it, depending on the sign of SMF, operates on a current placed in SMF. There is also observed a boomerang effect when on the ends of the conductor making longitudinal movement in SMF, EMF induction is induced. Generator of new type [ 9 ] operates on this principle.

Accordingly, there is a basis to consider Maxwell's electrodynamics as a restricted theory. Let's undertake an attempt to develop generalized electrodynamics on the basis of general properties of vector potential. Thus to some extent it is possible to base on analogies in the description of two components of a magnetic field. It is known, that the non-stationary vector magnetic field generates vortical electric field. By analogy: non-stationary SMF generates potential electric field. This hypothesis has the strict proof and is confirmed experimentally [9].

### 3. Theoretical analysis

Let us assume that in some volume $\tau$ non-uniform and non-stationary SMF $B^*(x,y,z,t)$ is created. Presence of sources and drains of a potential electric field in each point as follows from the put forward hypothesis, is characterized by a derivative $\dfrac{\partial B^*}{\partial t}$. In elementary volume $d\tau$ presence of sources and drains of an electric field is determined by product $\dfrac{\partial B^*}{\partial t}d\tau$ and for all allocated defined volume we have:

$$\Phi^* = \int_\tau \frac{\partial B^*}{\partial t} d\tau. \tag{20}$$

On the other hand the same value represents the flux of vector $\vec{E}$ through surface *S* limiting allocated volume:

$$\Phi^* = \int_S E_n dS. \tag{21}$$

Using Ghauss theorem, from (20) and (21) it is possible to write down:

$$\int_{\tau} \frac{\partial B^*}{\partial t} d\tau = \int_{\tau} div\vec{E} d\tau,$$

and this leads to the primary correlation presented without a conclusion in the monograph *of G.V. Nikolaev* [5-6]:

$$\frac{\partial B^*}{\partial t} = div\vec{E}. \tag{22}$$

Let's formulate the prototype law of electromagnetic induction in the differential form: ***a point of space in which the non-stationary scalar magnetic field has been created, is a source or a drain of an electric field.*** Consequently, one of the equations of generalized electrodynamics looks like:

$$div\vec{D} = \rho + \varepsilon'\varepsilon_0 \frac{\partial B^*}{\partial t}, \tag{23}$$

where $\vec{D}$ - vector of electric field induction, $\varepsilon'$ - relative dielectric permeability of a medium, $\rho$ - electric charge density.

As a result of the research carried out above it is possible to write down the full system of the electrodynamic equations that take into account the two constituents of magnetic fields (vortical and potential), as well as the two constituents of electric fields (vortical and potential). The known system of Maxwell's equations does not reflect this symmetry of electrodynamics and consequently is not full. The system of generalized electrodynamics looks as follows [4]:

$$rot\vec{H} + gradH^* = \vec{j} + \frac{\partial \vec{D}}{\partial t}, \tag{24}$$

$$rot\vec{E} = -\frac{\partial \vec{B}}{\partial t}, \tag{25}$$

$$div\vec{D} = \rho + \varepsilon'\varepsilon_0 \frac{\partial B^*}{\partial t}, \tag{26}$$

$$div\vec{H} = 0. \tag{27}$$

It appears from the equation (24), that conduction current $\vec{j}$ creates both vector and scalar magnetic field. Generally both these components of the one magnetic field are non-stationary. Due to the shift of an induction of vector magnetic field $\vec{B}$, as is known, the vortical electric field (25) is formed. The shift of SMF induction along with electric charges generates sources and drains of a potential electric field (26). Thus, the electric field generally includes potential and vortical components. Therefore when calculating derivative $\frac{\partial \vec{D}}{\partial t}$ there emerge vortical and potential (scalar) components of a magnetic field (25). That is displacement currents, as well as conduction currents, generate both components of a magnetic field: vector and scalar. The equation (27) specifies the absence of sources and drains of a vortical magnetic field. It is necessary to take on the Ohm law in the differential form to the equations (24) - (27). Ohm's law should be written down under condition of immovability of medium:

$$\vec{j} = \sigma\vec{E}, \tag{28}$$

where $\sigma$ - electroconductivity of medium, vector $\vec{E}$ - the sum of the intensities of both constituents of an electric field: vortical and potential.

The equations (24) - (28) are correct at the following assumptions:
1) all bodies present in an electromagnetic field are motionless,
2) units $\varepsilon', \mu', \sigma$ are the functions of coordinates and do not depend on time and the characteristics of an electromagnetic field.

It is not difficult to show, that the equation of indissolubility of generalized electrodynamics contains an additional member:

$$\frac{\partial \rho}{\partial t} + \varepsilon' \varepsilon_0 \frac{\partial^2 B^*}{\partial t^2} + div \vec{j} = 0. \qquad (29)$$

Thus in the point which is a source (drain) of an electric current there is a variable electric charge; and non-stationary SMF is necessarily created, besides $\frac{\partial^2 B^*}{\partial t^2} \neq 0$. Let's note that all the values in the differential equation (29) refer to one point of space. It is important to bear in mind when it is used together with the basic differential equations of electrodynamics (24)-(25) in which the variables in the right and left parts refer accordingly to various points of space located extremely close to each other.

In traditional electrodynamics, as is known, Lorenz antecedent is used:

$$div \vec{A} + \mu' \mu_0 \varepsilon \varepsilon_0 \frac{\partial \varphi}{\partial t} = 0, \qquad (30)$$

where $\varphi$ - scalar potential. It is introduced artificially and a physical meaning is not given to it. If we write down:

$$H^*(x', y', z', t) = -\frac{1}{\mu' \mu_0} div \vec{A} - \varepsilon \varepsilon_0 \frac{\partial \varphi}{\partial t}, \qquad (31)$$

then we get an opportunity to take into account a scalar component of the magnetic field and to check up its real existence. Let's note that in a stationary situation for vacuum (31) coincides with (8).

It is not difficult to show with the help of the equation (25), that in generalized electrodynamics expressing full intensity of an electric field by means of full electrodynamic and scalar potentials remain correct:

$$\vec{E} = -\frac{\partial \vec{A}}{\partial t} - grad \varphi. \qquad (32)$$

Having substituted expressions (12) and (31) in (24) with the account of (32) we shall receive:

$$\frac{1}{\mu' \mu_0} rot rot \vec{A}_O - \frac{1}{\mu' \mu_0} grad div \vec{A} - \varepsilon' \varepsilon_0 grad \frac{\partial \varphi}{\partial t} = j - \varepsilon' \varepsilon_0 \frac{\partial}{\partial t}\left(grad \varphi + \frac{\partial \vec{A}}{\partial t}\right).$$

We come to Dalamber equation for full vector potential:

$$\Delta \vec{A} - \varepsilon' \varepsilon_0 \mu' \mu_0 \frac{\partial^2 \vec{A}}{\partial t^2} = -\mu' \mu_0 \vec{j}. \qquad (33)$$

Similarly having transformed the equation (26) we shall receive the wave equation for scalar potential:

$$\Delta \varphi - \mu' \mu_0 \varepsilon' \varepsilon_0 \frac{\partial^2 \varphi}{\partial t^2} = -\frac{\rho}{\varepsilon' \varepsilon_0}. \qquad (34)$$

Consequently the process of radiation of electromagnetic waves in generalized electrodynamics, as well as in traditional one, is described by the two wave equations written down accordingly for vector and scalar potentials. But in contrast to the known limited theory these equations describe occurrence of an electromagnetic field at the account of a conduction current. The field consists of two magnetic components (vortical and potential), and also two electric components (vortical and potential). That is there is observed a certain symmetry between magnetic and electric fields that form one electromagnetic field. It is possible to tell that $\vec{A}$ and $\varphi$ form one four-dimensional vector – potential, whose properties describe all macroscopic electrodynamic phenomena.

At this phase of investigation it becomes clear that *the use of antecedent (2) in magnetostatics and Lorenz antecedent (30) in electrodynamics has resulted in restriction of the theory and exclusion from the consideration of the potential magnetic component, and, consequently, all phenomena connected to it.*

We shall receive the wave equations for an electromagnetic field with sources on the basis of the equations (24)-(27). Having applied operator $\partial/\partial t$ to the equation (24), taking into account (24) and (25), after transformations we have:

$$\Delta \vec{E} - \mu'\mu_0 \varepsilon'\varepsilon_0 \frac{\partial^2 \vec{E}}{\partial t^2} = \mu'\mu_0 \frac{\partial \vec{j}}{\partial t} + \frac{1}{\varepsilon'\varepsilon_0} grad\rho. \tag{35}$$

As the potential electric field is generated by electric charges, and the vortical field is connected to the locked-in electric currents, it is possible to split (35) into two independent equations:

$$\Delta \vec{E}_\rightarrow - \mu'\mu_0 \varepsilon'\varepsilon_0 \frac{\partial^2 \vec{E}_\rightarrow}{\partial t^2} = \frac{1}{\varepsilon'\varepsilon_0} grad\rho, \tag{36}$$

$$\Delta \vec{E}_O - \mu'\mu_0 \varepsilon'\varepsilon_0 \frac{\partial^2 \vec{E}_O}{\partial t^2} = \mu'\mu_0 \frac{\partial \vec{j}}{\partial t}. \tag{37}$$

Having calculated a time derivative from the equation (25) with provision for (24) we shall receive the wave equation for vector $\vec{H}$:

$$\Delta \vec{H} - \varepsilon'\varepsilon_0 \mu'\mu_0 \frac{\partial^2 \vec{H}}{\partial t^2} = -rot\vec{j}. \tag{38}$$

Similarly, having transformed (26) with provision for (24) we shall receive the wave equation for scalar function $H^*$:

$$\Delta H^* - \varepsilon'\varepsilon_0 \mu'\mu_0 \frac{\partial^2 H^*}{\partial t^2} = \frac{\partial \rho}{\partial t} + div\vec{j}. \tag{39}$$

That is, if in some point of the electroconductive medium there is a shift of an electric charge, this medium is a source or a drain of an electric field (current) which generates SMF in adjacent points of the environment. Let's note that in the given wave equation the values on the right part and at the left part refer to various points of space; therefore if we use the generalized equation of continuity (29), we receive from (39):

$$\Delta H^*(x',y',z',t) - \varepsilon'\varepsilon_0 \mu'\mu_0 \frac{\partial^2 H^*(x',y',z',t)}{\partial t^2} = -\varepsilon'\varepsilon_0 \frac{\partial^2 B^*(x,y,z,t)}{\partial t^2}. \tag{40}$$

The second member of the left part is not compensated by the right member, as the functions contained in them refer to various points of space. From (40) it follows: non-stationary SMF created in some point of the electroconductive medium with coordinates $x,y,z$ generates in same point sources (drains) of an electric current which results in the occurrence of non-stationary SMF in the neighboring point of space $(x', y', z')$.

Consequently, *in the general case two electromagnetic waves are formed: one of them is determined by vortical vectors $\vec{E}_O$ And $\vec{H}$, and another - by potential vector $\vec{E}_\rightarrow$ and scalar function $H^*$.* The first type of waves is known in electrodynamics for a long time, their properties are well researched. In particular it is shown, that far apart from the source of radiation (plane wave) vectors $\vec{E}_O$ and $\vec{H}$ are arranged perpendicularly to the direction of electromagnetic wave propagation; therefore these waves are referred to as *transversal.* The second type of waves is scarcely investigated, though there are a lot of publications about longitudinal electromagnetic waves. The term "longitudinal electromagnetic waves" in classical electrodynamics stands for some component of a non-plane wave determined by vortical vectors $\vec{E}_O$ and $\vec{H}$. Longitudinal constituents of usual electromagnetic waves are formed, for example, when it is propagated through conducting guides. This phenomenon does not exceed the limits of traditional electrodynamics. In originating generalized electrodynamics there is also the need to use the term '*longitudinal electromagnetic waves*' to define the second type of waves, however its meaning differs from the concept used in the traditional theory. *It is suggested to determine as 'longitudinal electromagnetic waves' the waves formed as a result of the shift of potential vector $\vec{E}_\rightarrow$ and scalar intensity $H^*$, and extending in the direction of vector $\vec{E}_\rightarrow$* (E-waves).

Let's note that the speeds of spreading transversal and longitudinal electromagnetic waves are identical:

$$V_\perp = V_\| = \frac{1}{\sqrt{\varepsilon'\varepsilon_0 \mu'\mu_0}} = \frac{c}{\sqrt{\varepsilon'\mu'_0}},$$

where $c = \frac{1}{\sqrt{\varepsilon_0 \mu_0}}$ - velocity of light in vacuum. Here lies indissoluble connection of all the constituents of the electromagnetic process and impossibility of their complete positional separation in the general case. Moreover, it is possible to show that a transversal electromagnetic wave derivates a longitudinal $E$-wave, and vice versa.

## 4. Discussion of the results

The theory of longitudinal electromagnetic waves demands separate consideration. Here we shall note only, that it is well coordinated to known experiments *of Nikolaev G.V.* [5-6] with two-planimetric aerials, one of which serves as oscillator, another - as the receiver of longitudinal electromagnetic waves.

This theory also explains the results of very exciting experiments described by *C. Monstein* and *J. P. Wesley* [10]. Their first experiment shows the transfer of energy at the account of longitudinal waves between the plates of the condenser which have been moved apart farther than the length of the wave. In addition the filter absorbing transversal waves is installed between the plates. In the second experiment there were used the spherical aerials installed from 10 to 1000 mm apart. Variable electric charge being created on the radiating aerial, the reception aerial registered a signal fading proportionally as the square of distance from the source of radiation. Such dependence is explained by the following: a non-directional radiating aerial was used. *C. Monstein* and *J. P. Wesley* do not use the concept of SMF and do not aim at going beyond classical electrodynamics, therefore their approach to a problem turned out unilateral: their explanation is based only on Poisson equation for scalar electric potential and they investigate only the shift of potential vector $\vec{E}_\rightarrow$. Nevertheless, C. Monstein and J. P. Wesley's experiments are extremely important for understanding of the researched phenomenon, as they show the limitation of the received view of the electromagnetic field and allow us to make a link to generalized electrodynamics.

A rather detailed analysis of the given problem is made by Enshin A.V. and Iliodorov V.A. [11-12]. The authors experimentally established "…that, when the paramagnetic gas medium is exposed to laser radiation with special spectral structure, there occurs the polarization of spins of molecules or atoms contained. Under the resonant influence of pondemotor forces of laser radiation there occurs the formation of quazicrystal spin-polarized structure with clear-cut ferromagnetic properties. That is in spin-polarized medium the appearance of macroscopical quantum effects is possible. In spin-polarized structure of paramagnetic gas takes place the transformation of laser radiation to longitudinal electromagnetic waves (research was carried out with the molecular gases with not compensated electronic or nuclear spin and some other substances). That is the waves where vector of an electric field coincides with the direction of a wave vector. Such transformation becomes possible because the radiation is accomplished not by separate electrons that really can radiate only transversal electromagnetic waves, but the aggregate of external electrons combined by exchange interaction and behaving as a quantum liquid. Longitudinal electromagnetic waves generated by spin-polarized structure have much higher orientation and coherence than the initial laser radiation. And we mean higher not by percent or times, but exponents. The section of absorption of longitudinal radiation is also much less than transversal."

As Enshin A.V. and Iliodorov V.A investigate the quantum phenomena in their works it's notable that the problem of longitudinal electromagnetic waves for the first time arose in quantum electrodynamics in the 30-s of the last century, since the 4-dimensional mathematical apparatus demanded their introduction. However, in order to satisfy Maxwell's theory special calibrating antecedents excluding longitudinal electromagnetic waves were introduced, and the longitudinal waves were declared "non-physical". Only in the 90-s this problem was risen again by Hvorostenko N.P. [13]. Since quantum electrodynamics operates with 4-dimensional vectors, vector $\vec{H}$ and some scalar function are used for the characteristic of a magnetic field; electric field is characterized by vector $\vec{E}$ and an additional scalar function. The last scalar function is characteristic only for quantum interaction (spinory electrodynamics), it is not necessary to be introduced when considering only macroscopical phenomena (conduction currents). Thus, the quantum theory is the most general case of generalized electrodynamics, it presents the complete symmetry in the description of an electromagnetic field. On the basis of quantum electrodynamics' apparatus Hvorostenko N.P. has shown that three are three possible modes of electromagnetic waves: one transversal and two longitudinal (*E*-waves and *H*-waves).

Four differential wave equations are received by Hvorostenko N.P: two of them are vector equations, two more - scalar. Three of these equations (taking into consideration difference in designations without the account of the quantum phenomena) coincide with the equations (35) received by us, (38) and (39). The fourth equation is specific to quantum electrodynamics and we haven't considered it.

Hvorostenko N.P. erroneously applied the usual equation of continuity (without an additional member) to (39), thus the right part of the equation (39) resulted in zero, and the author drew a conclusion on the absence of material sources stimulating longitudinal *E*-waves and declared them non-physical. The inaccuracy of this conclusion deprived all the further *E*-waves research of a theoretical basis; as a result, they have not been acknowledged by the official science until now, despite the experimental facts confirming their real existence. The longitudinal *H*-waves are acknowledged by the quantum theory, they are created by spin-magnetic axial currents.

The results of research of Enshin A.V. and Iliodorov V.A. are apparently explained on the basis of a complex of the phenomena of all the three types of electromagnetic waves. Transversal laser radiation results in establishing in the substance of electronic quantum structures of toroidal type with a certain orientation of their axes. These structures, as we showed, are capable of generating longitudinal *E*-waves. At the quantum level the formation of *H*-waves is possible, they are compensated on macrolevel.

## 5. Conclusion

Relying on the results of the above mentioned publications and our research, it is possible to draw the conclusion that all three types of electromagnetic waves are physically real, moreover in the most general case all of them are interconnected, derivate each other forming one electromagnetic process. Theoretical and experimental study of properties of longitudinal electromagnetic *E*-*waves* will allow us to understand the essence of the electromagnetic phenomena deeper and to substantially expand the area of their practical use. Generalized electrodynamics (the macroscopical theory) which was developed with the account of all the properties of 4-dimensional electrodynamic potential is capable to explain numerous paradoxes of modern electrodynamics and to take a new look at the fundamental concepts of physics.